\documentclass[10pt,letterpaper]{article}
\usepackage[top=1in,left=1in,right=1in, footskip=0.75in,marginparwidth=1in]{geometry}

\usepackage[utf8]{inputenc}

\usepackage{cite}

\usepackage{nameref,hyperref}

\usepackage[right]{lineno}

\usepackage{microtype}
\DisableLigatures[f]{encoding = *, family = * }

\usepackage{changepage}

\usepackage[aboveskip=1pt,labelfont=bf,labelsep=period,singlelinecheck=off]{caption}

\makeatletter
\renewcommand{\@biblabel}[1]{\quad#1.}
\makeatother

\usepackage{lastpage,fancyhdr,graphicx}
\usepackage{epstopdf}
\fancyheadoffset[L]{1in}
\fancyfootoffset[L]{1in}

\usepackage{color}

\definecolor{Gray}{gray}{.25}

\usepackage{graphicx}

\usepackage{sidecap}

\usepackage{wrapfig}
\usepackage[pscoord]{eso-pic}
\usepackage[fulladjust]{marginnote}
\reversemarginpar

\begin{document}
\vspace*{0.35in}

\begin{flushleft}
{\Large
\textbf\newline{Spin polarized electron beams production beyond III-V semiconductors}
}
\newline
\\
L. Cultrera\textsuperscript{1,*}
\\
\bigskip
\bf{1} Brookhaven National Laboratory, Upton, NY 11973
\\
\bigskip
* lcultrera@bnl.gov

\end{flushleft}

\section*{Abstract}
This paper summarizes the state of the art of photocathode based on III-V semiconductors for spin polarized electron beam production. The limitations preventing this class of material to provide the long term reliability at the highest average beam currents necessary for some of the new accelerator facilities or proposed upgrades of existing ones are illustrated. Promising alternative classes of materials are identified showing properties that can be leveraged to synthesize photocathode structures that can outperform III-V semiconductors in the production of spin polarized electron beams and support the operating conditions of advanced electron sources for new facilities.


\section*{Introduction}
The use of spin polarized electron source finds application in a wide range of electron accelerators of interest for HEP. The planned International Linear Collider \cite{1} is designed to collide spin polarized electrons and positrons at very large energies in the TeV scale, SuperKEKB is considering upgrading to a highly spin polarized electron source \cite{2}. These facilities are projected to require relatively modest amount of average current, not more that few hundreds of $\mu$A that can be produced by state-of-the-art photoelectron sources equipped with GaAs based photocathodes. On another hand, other facilities like the Large Hadron electron Collider \cite{3} are planning to operate with very high average electron beam currents, tens of mA, which are well beyond the current state-of-the-art. Furthermore, the need of developing sources capable of providing highly spin polarized electron with high average beam currents arises from the recent demonstration of efficient transfer of the high spin polarization from electrons to positrons via a two-step process: in the first step polarized bremsstrahlung radiation is generated by a polarized electron beam in a high-Z target; in the second step this polarized bremsstrahlung radiation produces polarized positrons by pair-production process in the same target \cite{4}. Given the relative low efficiency of the process of production and subsequent capture of the positron beam (~10$^{-4}$-10$^{-3}$) there is the need of developing electron sources that can produce tens of mA of average spin polarized beam current to support the production of polarized positrons in the $\mu$A range.

\section*{III-Vs photocathodes for spin polarized electron beams}

When it comes to the production of spin polarized electrons beam in a photoelectron gun there are currently no other options in photocathode technology that do not make use of GaAs-based photocathodes structures. The discovery of spin polarized photoemission from GaAs was first reported in 1975 for bulk GaAs \cite{5}. The electron beam spin polarization observed in bulk GaAs comes from the different probabilities associated to the optical transition from the energy degenerate heavy holes and light holes valence bands when circularly polarized light is used to excite photoelectrons. In bulk GaAs, large Quantum Efficiencies (QE) are possible (typically 0.1 electrons per number of incident photons) with Electron Spin Polarization (ESP) theoretically limited to a maximum value of 50\% (but typically ESP achieved values in bulk GaAs are on the range of 40\% limited by several depolarization mechanisms during photoelectrons' diffusion in the bulk).

In 1992, the use of a strained GaAs pseudo-morphically grown on a relaxed GaAsP layer allowed breaking the energy degeneracy of the valence bands. With the removal of the degeneracy an energy selective excitation of photoelectrons with a single spin polarization was demonstrated. Photoelectrons achieved ESP larger than 80\%, but with a typical QE of the process in 10$^{-3}$ range due to the limited thickness of the GaAs strained layer \cite{6}.

The use of strained superlattices (SL) based on GaAs/GaAsP allowed by leveraging the band engineering of quantum wells structures to increase the separation between heavy hole and light hole bands \cite{7} and achieve even larger spin polarization (above 85\%) and QE (just above 1\%). Advanced SL design techniques made use of the strain compensation technique to decrease the defect density due to the accumulated strain in the SL. The design of these structures is based on alternating layers with tensile and compressive strain so that on average the SL results un-strained \cite{8}. By minimizing the amount of defects in the crystal lattice this design enabled further increase of the spin polarization (up to 92\%) with an achieved QE just above 2\%. Even if SL strain compensated structures have been demonstrated with very large number of quantum wells (up to 90 pairs) a simultaneous maximization of the QE and ESP is hard to achieve. As QE increases, because of the larger thickness and optical absorption in the SL medium, the ESP decreases because a larger fraction of electrons diffusing from the deepest layers is affected by spin depolarization mechanisms during the transport before reaching the interface to be extracted into vacuum.

A Distributed Bragg Reflector (DBR) grown underneath the SL lattice with the help of an intermediate layer providing the right optical thickness separating the DBR and SL can be used to design and realize a Fabry Perot resonator that effectively trap the light in a relatively narrow bandwidth inside the photocathode structure. Photons will be reflected back and forth between the DBR and the SL outer layers allowing to effectively increase the light absorption and as
consequence the QE. One of such a structure demonstrated record performances achieving at the same wavelength (775 nm) an ESP of 84\% and a QE of 6.4\% \cite{9}. As the number of layers and materials increases the growth of such structures now becomes quite complicated (more than 50 layers and 5 different materials are often required) affecting the rate of success in obtaining the design specification.

\subsection*{Negative Electron Affinity and operational lifetime
}

The electronic band gap of GaAs is about 1.4 eV and with surface electron affinity of 3-3.5 eV reaching the threshold of electrons photoemission requires UV light. To extract electrons that have been excited with photon energies close to band gap energy it is necessary to modify the material surface electron affinity so that the vacuum level will lay below the conduction band minimum. This condition is knows as Negative Electron Affinity (NEA). Such a state is achieved by exposing the clean GaAs surface to Cs vapors allowing the formation of strong electric dipole field near the surface. The intense electric field strongly reduces the height and width of the potential barrier and permits the extraction of electron already relaxed to the bottom of the conduction band. Generating such strong dipole and achieving the NEA condition usually require less than a mono-layer of Cs on the surface. This was simultaneously confirmed by experimental data and numerical computations of work function variation as function of the Cs exposure as well as from advanced surface imaging technique on NEA activated GaAs \cite{10,11}. The strength of the surface dipole is usually enhanced by using alternating layers of Cs and O$_2$ or NF$_3$ gasses. Unfortunately, due to the extreme reactivity of these activating layers the vacuum requirements necessary to sustain the NEA condition are quite challenging: in order to have a sufficiently long lifetime to be of any practical use the GaAs surface activated to NEA are usually operated in vacuum levels below 10$^{-11}$ Torr.
Such drastic vacuum requirements have so far limited the operation of GaAs based photocathodes only on High Voltage DC electron guns. In these electron guns the vacuum vessel is usually quite large to prevent electric discharges between the cathode stalk which is held at high voltage and the surrounding UHV chamber walls at ground potential. The large volume of the gun vessel can host a large amount of Non Evaporable Getter pump providing very large pumping speeds (many thousands of liter per second) allowing to reach vacuum levels of 10$^{-12}$ Torr or even better \cite{12,13,14}.

The dark lifetime measures the 1/e decay of the QE when the cathode is idling in the gun vessel with no electric field and no charge extraction. In a typical DC gun designed to reach few 10$^{-12}$ Torr of base pressure the dark lifetime can easily reach few thousands of hours \cite{13}. In such extreme vacuum and in the absence of laser illumination and electric field allowing for electron extraction, the only mechanisms responsible for the degradation of the efficiency is the chemical poisoning of the activation layer which happens in response to the exposure of gas molecules in the residual gas: even in the extreme vacuum achieved in DC gun small quantities of oxygen containing species are still present and for some of them, like O$_2$ and CO$_2$, only a fraction of a monolayer is sufficient to produce dramatic reduction of photoemission \cite{15}.

When photocathode are used to generate an electron beam a better metric to define the lifetime is perhaps the charge lifetime which measure the 1/e decay of their quantum efficiency as function of the extracted integrated charge. In this case the QE degradation can be affected not only by the chemical poisoning, but also by the chemical desorption of alkali metals induced by the thermal laser heating of the photocathode surface \cite{16} and by the back bombardment of ions accelerated by the field existing in the anode-cathode gap and striking at the photocathode surface \cite{17}. The number of experimental parameters influencing the scaling laws associated to these last two degrading processes and the way they are entwined complicates the scenario and it is often difficult comparing results obtained with different electron guns or distinct experimental conditions. Just to make an example: the ions in in the cathode anode gap are usually produced by ionization of the residual gas molecules by electron impact from the primary electron beam. The number of ions will be proportional to the gas density in that region that may be influenced by electron induced desorption from stray energetic electrons hitting the surrounding surfaces which are difficult to predict. At average beam currents levels of 1 mA and 4 mA charge lifetimes of 200 C and 80C have been reported \cite{18,19} corresponding to a 1/e lifetime during operation of about 2.5 days and 5 hours respectively. Taking the most favorable case (200 C of charge lifetime) and scaling it for an average current of 20 mA (like the one required by the ERL proposed in LHeC) will result in a operating 1/e lifetime of less than 3 hours. These estimates seem also quite well aligned with the operational lifetime measured at the DC gun of Cornell University where GaAs operated at record average currents up to 50 mA showed 1/e QE lifetimes of about 1 hour \cite{20}. 

The advanced GaAs based cathodes made with SL-DBR structure offer some interesting perspectives to increase the operational lifetime because given the large initial achievable QE they require less laser power to produce the design beam current leaving larger margins to compensate for QE degradation, but they are still burdened by the complex growth process which make them difficult to synthesize and and achieve the design parameters.
In order to produce spin polarized electrons with beam parameters beyond the current state of the art, especially in terms of average current, and enable the operation of photocathodes in high gradient RF electron guns, to leverage higher achievable beam brightness, it is necessary to develop robust alternatives to GaAs and III-Vs technology. Other classes of materials have reached in the last decades the technological maturity that warrant them to be studied and possibly employed as alternative electron sources for spin polarized beams. Other emerging materials and schemes for spin polarized electron generation and transport, showing promising result obtained in other disciplines, especially in spintronics, can be investigated and an increased collaborations between the communities of accelerator physicist and materials scientist must be promoted to enable further discoveries. Finallay, the need of electron gun test facilities capable of testing these materials in a wider range of operating conditions should be considered a indispensable complementary effort necessary to validate results obtained in a laboratory setting typical of material science experiments.

\section*{III-materials}

The technology of III-Nitride materials (GaN, AlN, InN and their ternary alloys) has seen in the last decade several advances largely driven by industry for their application in Ligth Emitting Diodes and high-power microelectronic devices. These materials are extremely robust and chemically stable also a very high temperature (GaN start to thermally decomposes at temperatures larger than 700 C). Some of their photoemission properties are already well known because they have been studied as possible photoemitters in solar blind photon detector \cite{21}. These family of photocathode materials present a wide range of interesting and unique properties: among them their band gap tunability from UV to IR that can be achieved by tailoring the composition of ternary alloys of GaN, InN, AlN (see figure 1). The plots in figure 1 shows that the use of ternary alloys of III-Nitrides allows in principle electron photoexcitation with lower photon energies enabling the operation of such photocathode with laser system in the visible or near UV spectral range that can be transported and manipulated by optic systems characterized by low losses.
Quite relevant for application in electron beams generation is the demonstration that by leveraging the polarization charges typical of III-Nitride material in their stable wurtzite structure, layered photocathodes can be designed to achieve a NEA condition without the use of Cs at the surface \cite{22}.

\begin{figure} 
\begin{center}
    \includegraphics[width=3.75in]{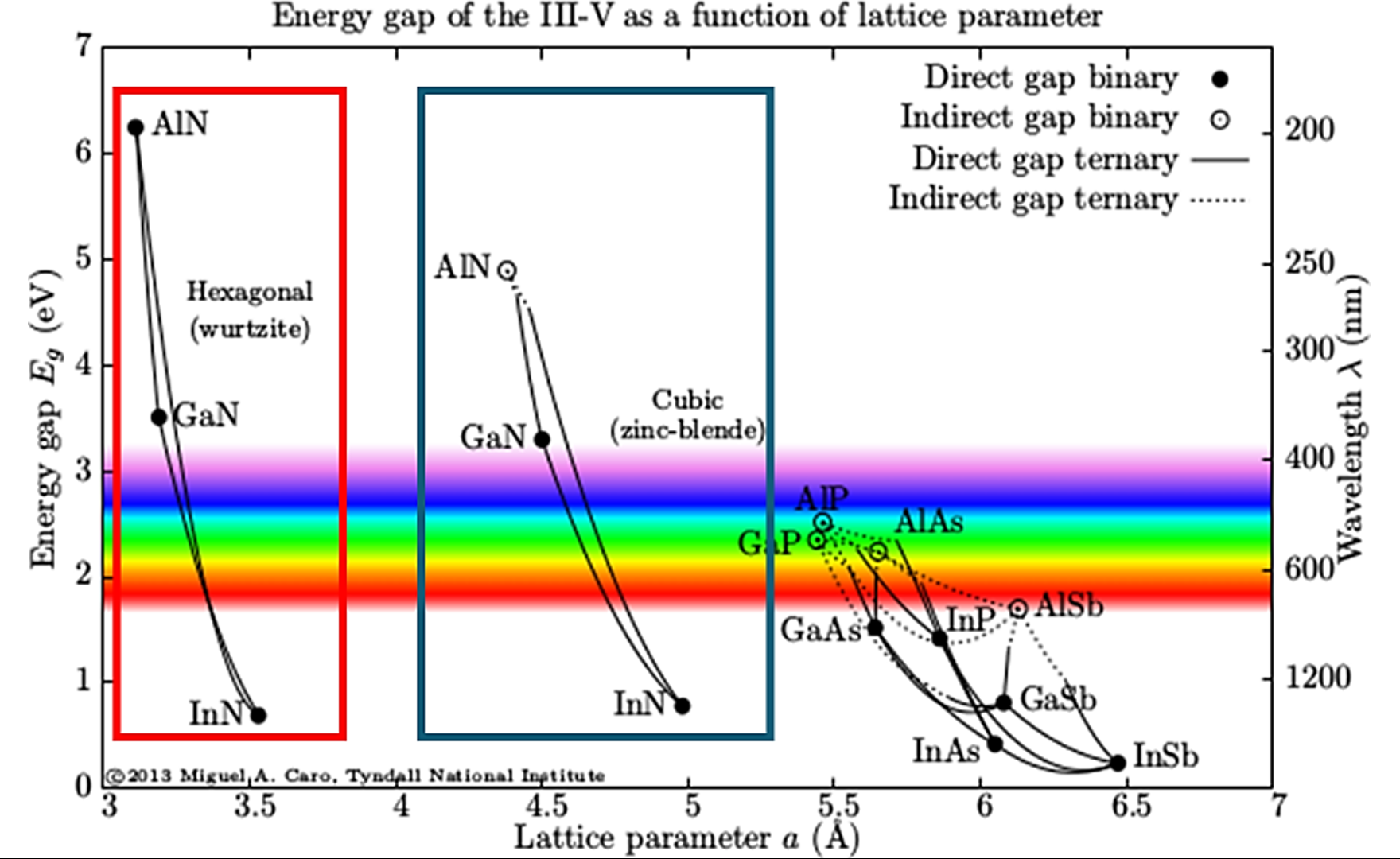}
\end{center}
\caption{\color{Gray} \textbf{III-Nitride family of semiconductors can be synthesized in two different crystal structure: wurtzite and zinc-blende. Band-gap values and crystal lattice parameter are reported as function of the ternary alloy composition.}}
\label{fig1} 
\end{figure}
 
If the photocathode surface no longer needs to rely on Cs based layer for the formation of NEA and due to the high chemical stability of the III-nitrides materials the extreme vacuum requirements are not anymore necessary such photocathodes can in principle be operated also in high gradient RF and SRF guns. 
Preliminary measurements of the photoemission properties in terms of QE, MTE and lifetime were performed on a GaN N-polar photocathode \cite{23}. This studied structure was originally developed for UV photon detection rather than for electron beam production and showed high values of peak QEs (over 25\%), Mean Transverse Energies (MTEs) of photoelectrons comparable if not better than that of other semiconductor photocathodes (around 50 meV at near threshold) with a robustness such that no QE decay was measured over more than 24 hours of operation \cite{23}. This photocathode structure allowed the following results to be obtained at the operating wavelength of 265 nm: QE of 1x10$^{-3}$, MTE of about 100 meV and no observable decay of the QE over a span of more than 24 hours. After a transport in open air (using standard shipping via mail) the sample preparation only required a quick chemical etching in HCl – few seconds - and a bake at 500 C for a couple of hours. 

Other results supports the pictures that while the wurtzite GaN and III-Nitride in general present larger than GaAs spin de-polarization rates the electrons' spin lifetime can be larger than 200 ps at the cryogenic temperature of about 80 K (figure 2). \cite{24}.  
\begin{figure} 
\begin{center}
    \includegraphics[width=3.75in]{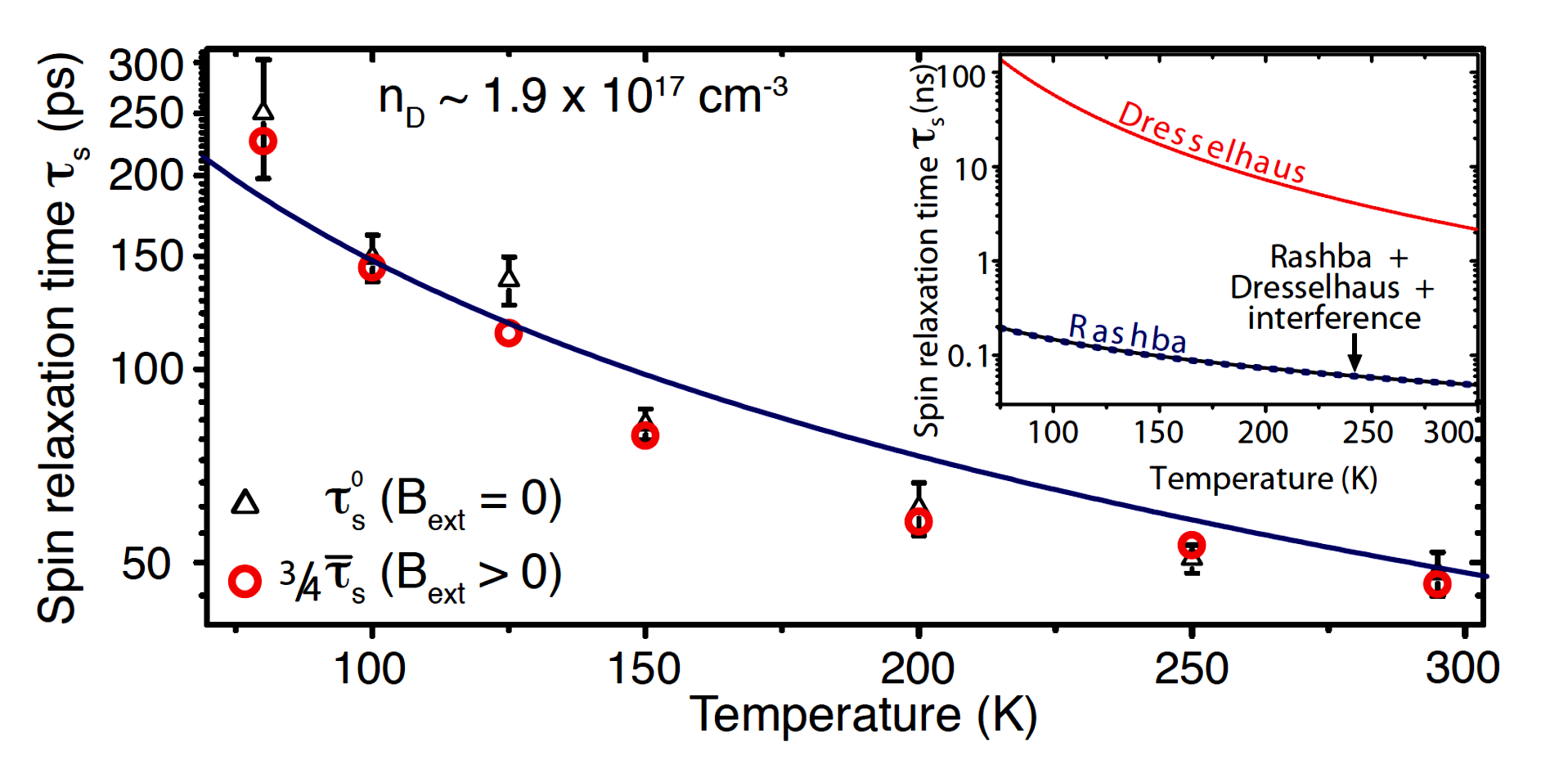}
\end{center}
\caption{\color{Gray} \textbf{Spin relaxation times for GaN as function of the temperature (reproduced from Ref.\cite{24}}
}
\end{figure}
With a GaN photocathode operated at cryogenic temperature in a moderate gradient like the one provided by a 3 and ½ cell SRF gun \cite{25} electron bunches with few 100s of pC can be extracted with laser pulses duration of ~5ps rms duration. Even with a few hundreds of ps in spin lifetime, if the cathode can be operated in the high electric fields typical of RF structures -where GaAs operation would be impossible- electron bunches with high spin polarization might be produced with bunch charges in the nC range. Due to the largest electric field at which RF guns operate if compared to DC guns, electron bunches can be extracted into vacuum in a time much shorter than the spin relaxation times and because of that it should be possible to preserve the initial spin polarization in the population of emitted electrons. Spin relaxation rates can be further decreased by lowering even more the photocathode operational temperature and also by increasing the crystalline quality of the material. By mitigating the presence of impurities and defects in the material it should be expected an increment of the mean free path of photoelectrons induced by a decrease of the associated scattering rates. The scattering mechanisms will not only drastically reduce the QE but may also randomize the spin orientation before the electron emission. 
This improvement in the quality of the crystal structure can be achieved leveraging single crystal substrates based on conductive GaN that have only been recently made commercially available that offer a native substrate for developing multilayered structures \cite{26}. 

In the less studied zinc-blende crystal structure GaN’s electron spin relaxation times have been measured to be as long as 5 ns: this value is even larger than the one measured for GaAs under similar conditions (3.6 ns) \cite{27} hence opening the possible scenario of developing photocathode based on III-Nitride that can provide spin polarized electron beam with longer electron bunch duration. NEA condition of III-nitride can be achieved only with Cs and it is more stable with respect to chemical poisoning than the one achieved on III-Vs \cite{27b}. 
Similar to what has been demonstrated for III-Vs with Cs-Sb-O and Cs-Te-O \cite{28,29,30,31} the use of heterostructures based on wide band gap materials might offer the possibility of providing an alternative solution for even more robust NEA activation layers.

\section*{Other systems}

Whenever in presence of a symmetric crystal structure a non-isotropic strain will results in a removal of the degeneracy between heavy holes and light holes valence band, if the atomic masses of the elements in the materials are large then the spin-orbit coupling will produce a large enough energy difference also in the split-off band with a potential to generate spin polarized electron beam population in the conduction band.The relatively large masses of Cd, Se, Zn and Te makes the II-VI an interesting candidate class of materials for the production of spin polarized electron beams.
\begin{figure} 
\begin{center}
    \includegraphics[width=3.75in]{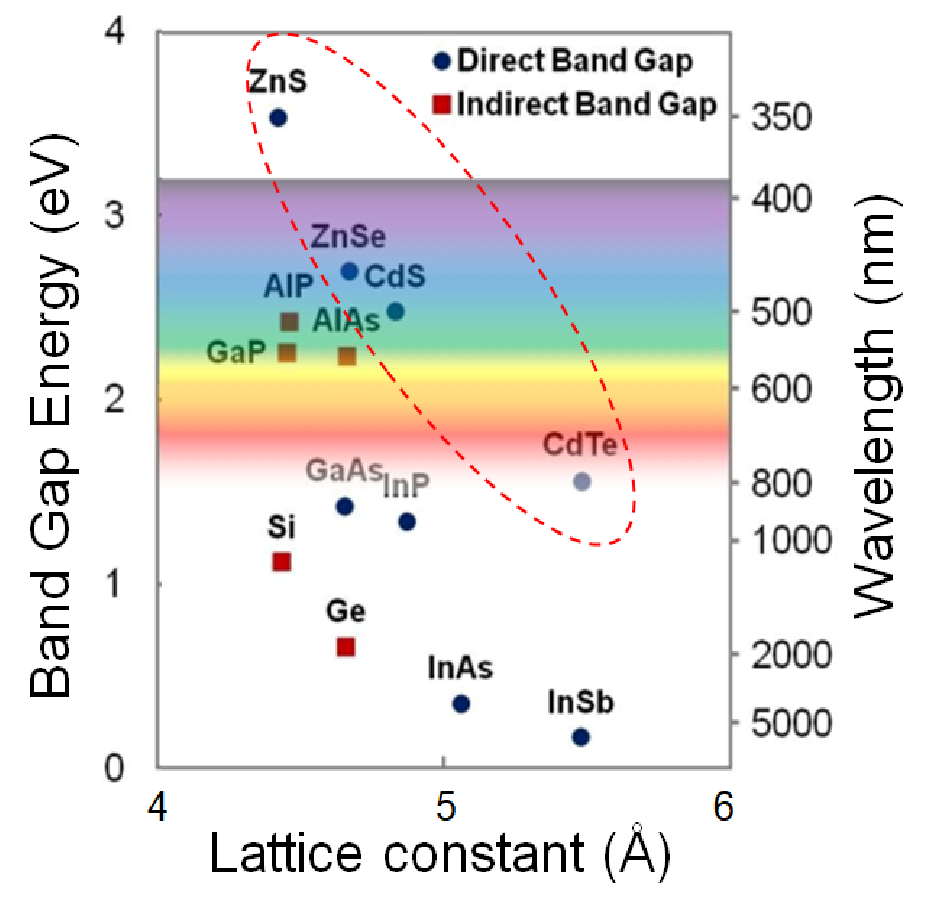}
\end{center}
\caption{\color{Gray} \textbf{Electronic band gap and lattice constant of common II-VI semiconductors.}}
\end{figure}
Several II-VI binary semiconductors (CdSe, ZnTe, CdTe…) and their ternary alloys which have been studied and developed for application photovoltaic high efficiency panels have band gap values ranging from UV to IR (see figure 3) and may be suitable as possible electron source for accelerator. When operated in photovoltaic structures II-VI semiconductors have demonstrated high external quantum efficiencies and have been grown also on transparent substrates \cite{32}. The first aspect is important as it indicates that photo-excited carriers lifetime is long enough to allow their transport across photovoltaic structures before recombining. The already demonstrated compatibility to transparent substrates can be leveraged to produce transmission mode photocathodes back illuminated \cite{33}. The II-VI materials should be studied initially as binary semiconductors to determine their possible use as electron sources and to understand if the possibility exists of activating them to NEA conditions using Cs in a way all similar to the one used in III-V semiconductors. Later, if initial test are promising, layered and strained structures based on the ternary alloys can be designed and grown and attempts at leveraging the internal fields, like in solar cells p-n junction, can be carried out to enhance the quantum efficiency of the photocathodes by accelerating the electrons towards the vacuum interface \cite{34}.

Spin-filter materials have been extensively studied in spintronics to form tunneling barrier where the heights of the barrier depend on the electron spin orientation allowing injection of polarized current in electronic devices (see figure 4). 
\begin{figure} 
\begin{center}
    \includegraphics[width=3.75in]{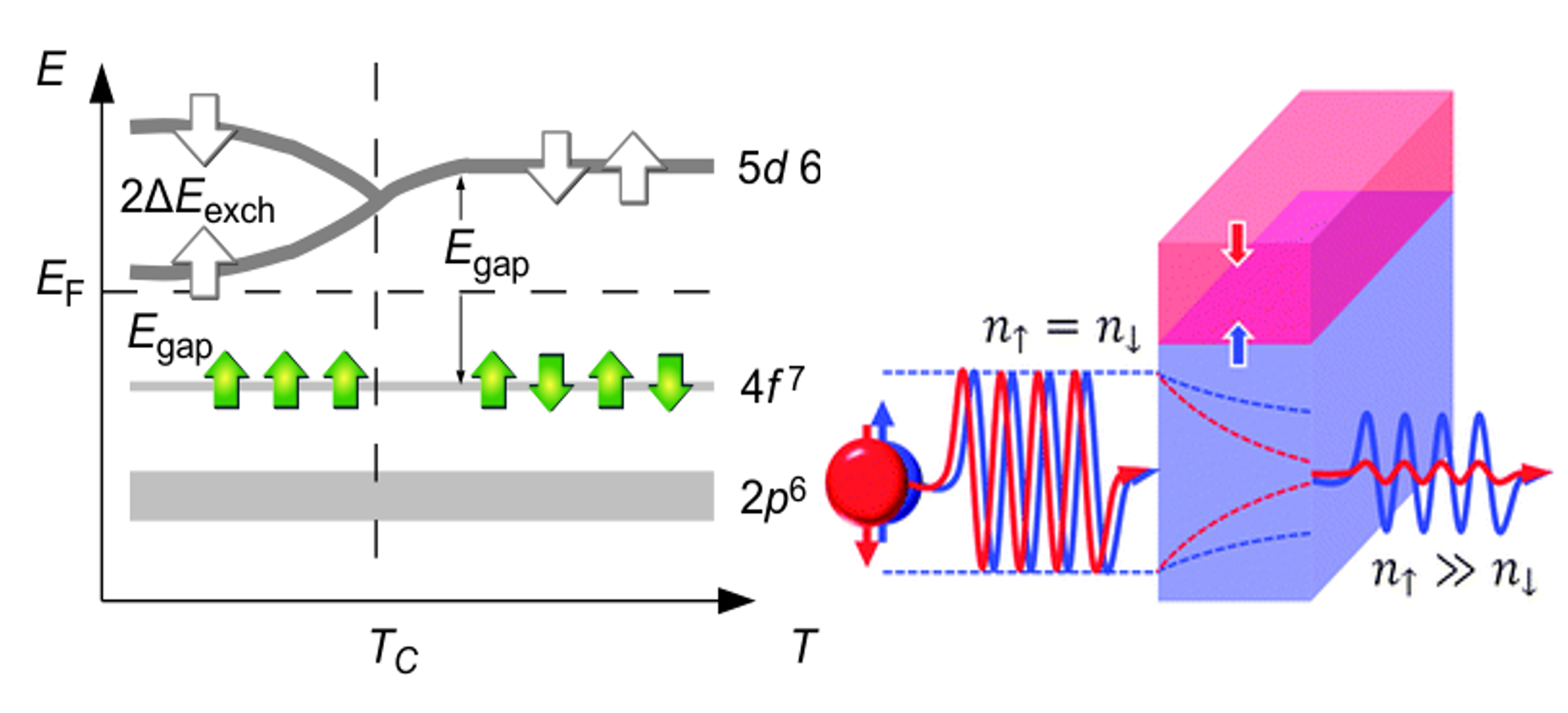}
\end{center}
\caption{\color{Gray} \textbf{Schematics of the spin filtering materials working principle: left) if the electron spins in the 4f levels are aligned (that is possible if temperature is $<T_c$) a spin dependent splitting of the bands can be induced with a magnitude equal to twice exchange splitting energy; right) the transmission coefficient for electrons tunneling through a barrier whose height is now spin dependent and can be used to generate highly polarized current (reproduced from Ref.\cite{34b}).}}
\end{figure}
Most of these materials require to be operated at cryogenic temperature to preserve the magnetization of electrons in the 4f orbitals that is at the origin of the band splitting and that allows controlling the spin direction in the tunneling currents. The spin-filter barrier made of EuS has already demonstrated the generation of highly polarized electron beams (89±7\%) during field emission experiments from a tungsten tip, but in order to do so it had to be cooled down to about 20 K, close to its Curie temperature \cite{35}. Other materials, like the spin filter EuO are extremely promising because the larger exchange splitting energy (0.54 eV vs. 0.36 eV of EuS \cite{36}) and the higher Curie temperature than the EuS (69 K instead of 16.6 K  for EuS \cite{37}) and indeed full spin polarization has been predicted for tunneling junctions based on EuO \cite{38}.
Amongst the half-metals, one of the possible candidates to perform photoemission experiments is the chromium dioxide. The predicted spin dependent density of states and work functions are reported in figure 4 along with photoelectron spin polarization measured using VUV photons \cite{39,40,41}. 
\begin{figure} [ht] 
\begin{center}
    \includegraphics[width=3.75in]{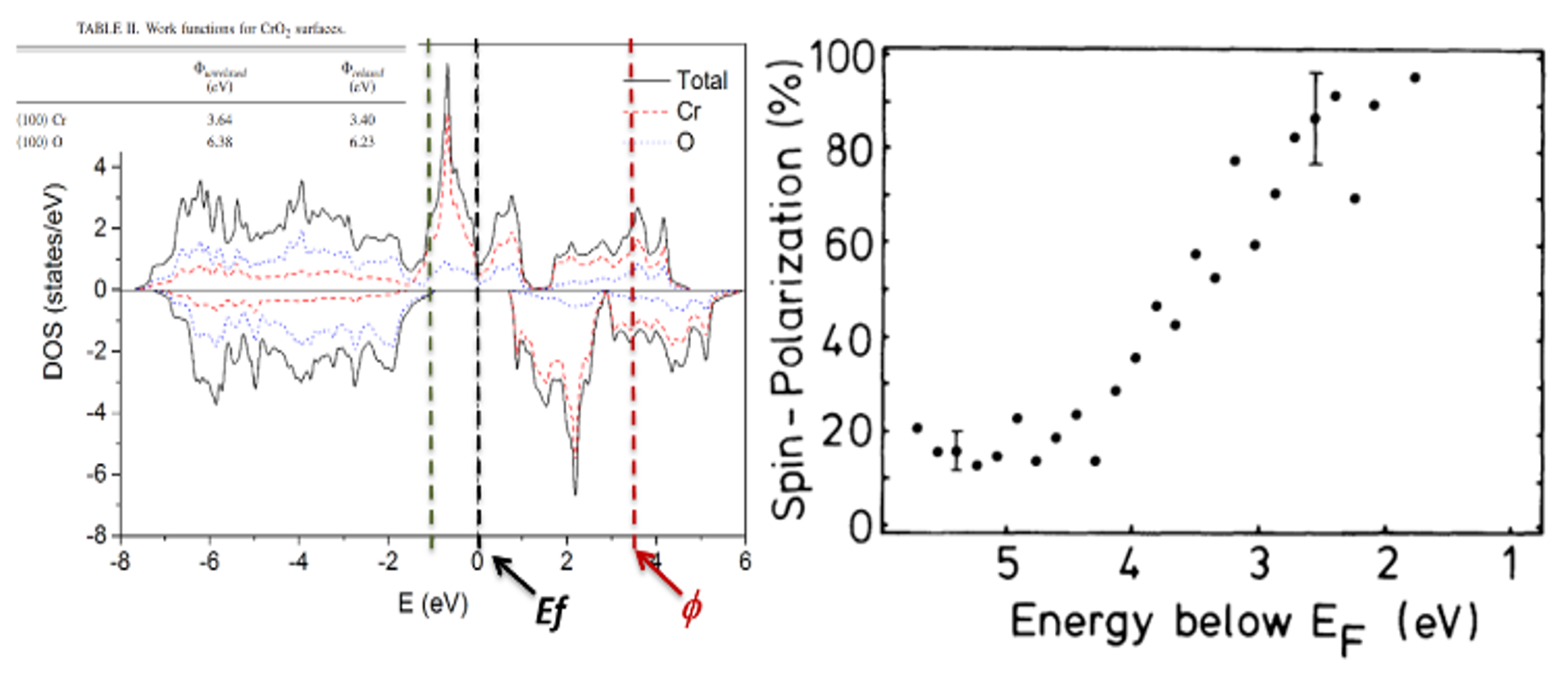}
\end{center}
\caption{\color{Gray} \textbf{Predicted work function, spin resolved density of states and experimentally measured photoelectrons’ spin polarization for chromium dioxide (reproduced from Ref.\cite{39,41}).}}
\end{figure}
CrO$_2$ can by synthesized using several different growth techniques \cite{42,43}. While the surface of CrO$_2$ is not as stable and energetically favored as the Cr$_2$O$_3$ a stable CrO$_2$ surface can be usually achieved or recovered by annealing the sample in a relatively high pressure of oxygen \cite{40}. An interesting property of the CrO$_2$ is that the reported Curie temperature of the material is relatively high (118 C) making these materials interesting for exploring its spin polarized photoemission properties already at room temperature and with UV light\cite{44}.

\section*{Extending the Space Of Operational Conditions}
Recent developments in laser technology have made the Optical Parametric Amplifiers (OPA) and Supercontinuum laser systems commercially available. These light sources offer to experimenters the possibility to continuously tune the laser wavelength from the UV to the IR. Furthermore, recent advances in electron gun technology have demonstrated electron guns based on DC technology \cite{45} and Superconducting RF technology \cite{25} capable of operating with the photocathode held at cryogenic temperatures. The possibility of expanding the range of wavelengths available for operating a photocathode in an electron gun with the capability of cooling the material down to cryogenic temperatures are two very important operational conditions that can be leveraged to improve the photoemission performance of photocathode materials. 
It is known that in the case of GaAs and GaN based photocathode for spin polarized electron beam production the operation at cryogenic temperatures can be leveraged to promote a mitigation of the processes leading to spin depolarization and push spin polarization close to theoretical limits in the extracted electron beams (see figure 6) \cite{46}.
\begin{figure} [ht] 
\begin{center}
    \includegraphics[width=3.75in]{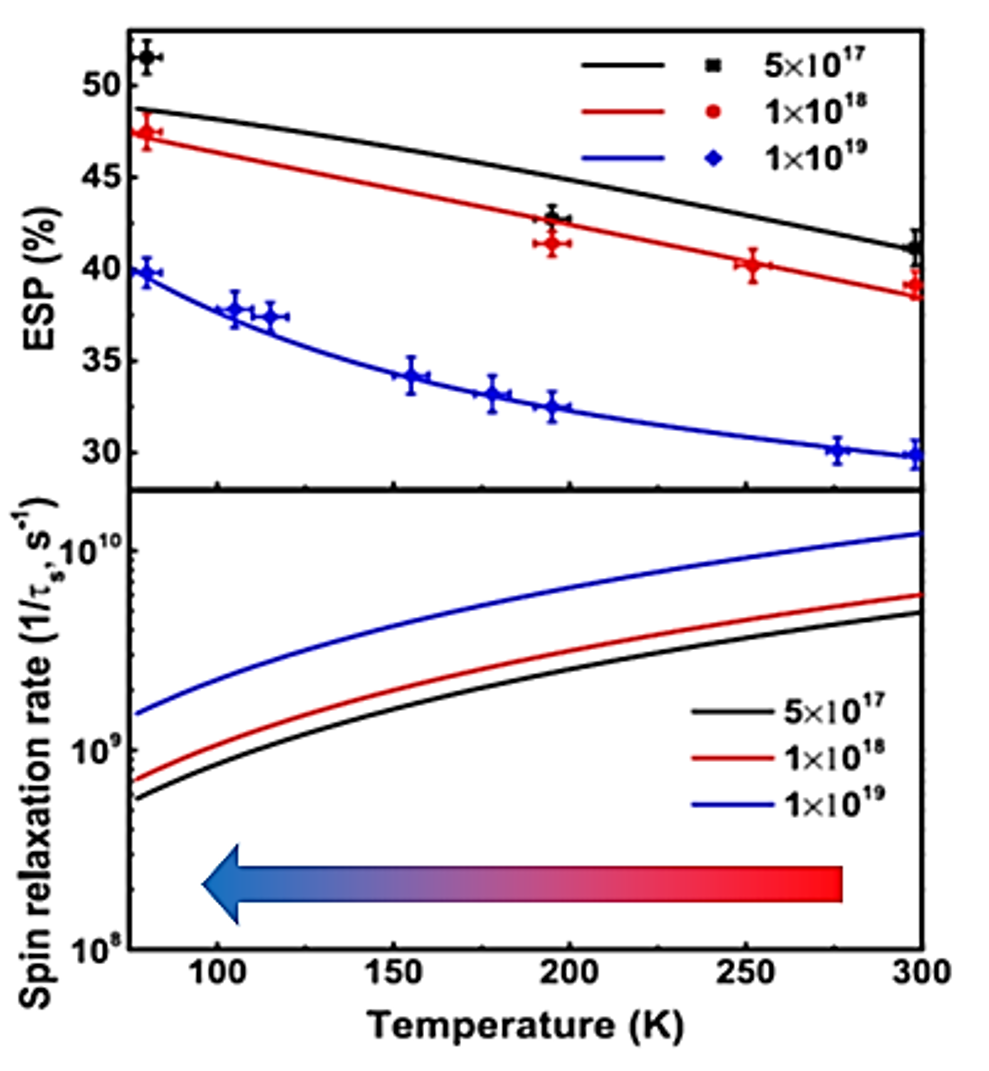}
\end{center}
\caption{\color{Gray} \textbf{As the temperature of the GaAs is lowered the spin depolarization rates of electron inside the materials decreases and the polarization of extracted electron increases up to the theoretical value of 50\% (reproduced from Ref.\cite{46}.}}
\end{figure}
Demonstrating in an electron gun that cryocooling can be used to mitigate spin polarization losses in robust heterostructure based photocathodes, could pave the way to high average current electron beam with high spin polarization which are required for a number of application (LHeC, PEPPo-II and future circular colliders based on high average current Energy Recovery Linacs).
The simultaneous use of cryogenic temperatures and drive laser wavelength tuning will enable access to a larger number of photocathode materials and also to the production of electron beam with reduced mean transverse energy (see figure 7). Experimental results have already demonstrated that one order of magnitude reduction in the MTEs of electrons can be obtained from metal \cite{47,48} and semiconductors \cite{49} once the emission wavelength is tuned close to the onset of photoemission. Exploring these new regimes of operation with new potential candidate materials is of great relevance and interest for producing beams with increased brightness and for the next generation of photoinjectors based on SRF guns planning to be operated with photocathodes held at cryogenic temperatures. 
The wide spectral range offered by new laser systems based on supercontinuum generation and OPA allow to overcome the limits imposed by the integer harmonic conversion of solid state lasers and open the photocathode research to explore a much larger range of materials. With a wavelength tuning capability ranging from 200 nm to 2000 nm a wide range of material (and an almost infinite number of layered structures built with them) become possible candidates to be operated in a photoinjector (see figure 1 and 3). 
\begin{figure} [ht] 
\begin{center}
    \includegraphics[width=3.75in]{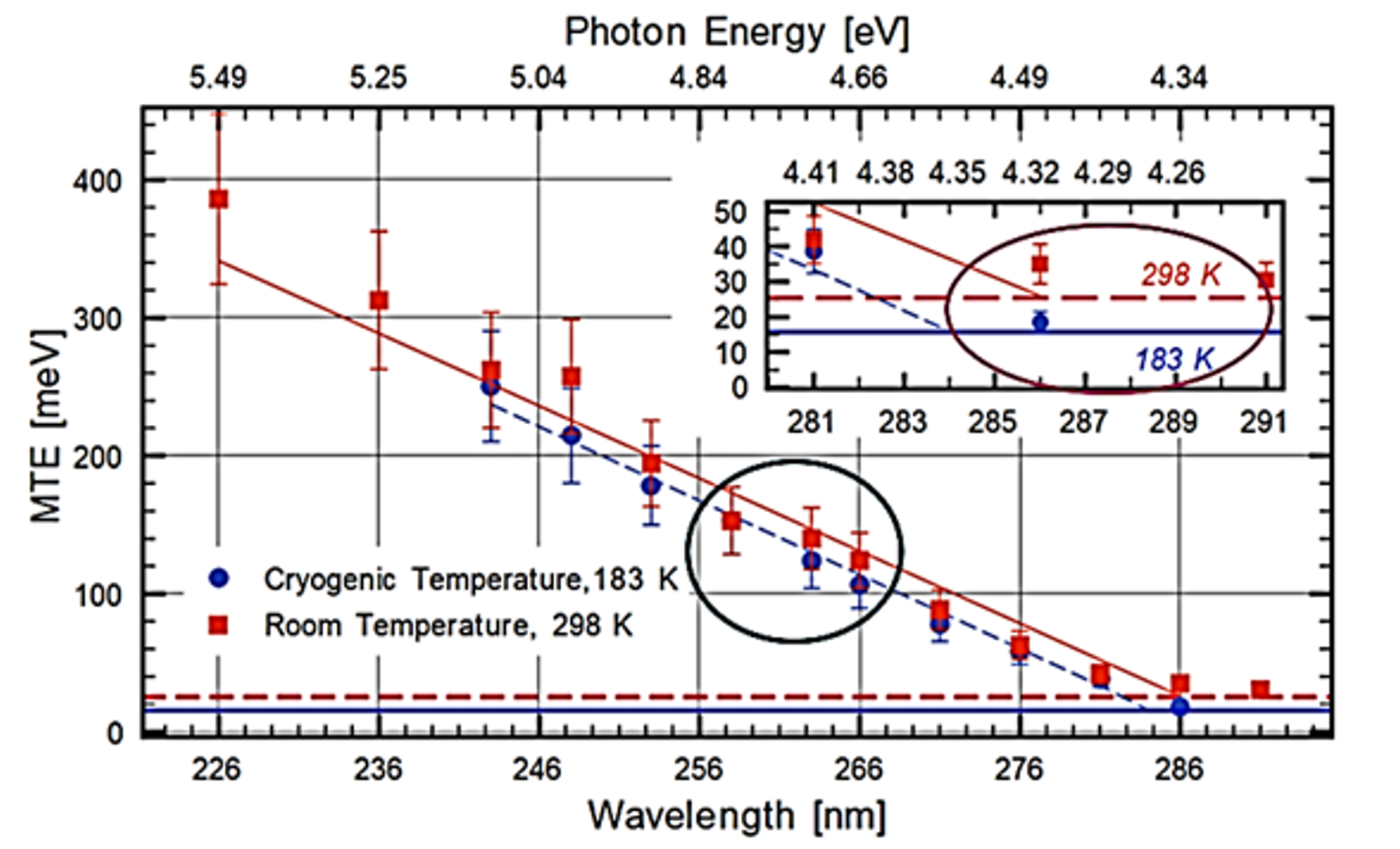}
\end{center}
\caption{\color{Gray} \textbf{The Mean Transverse Energy of photoelectron extracted from Cu(100) single crystal as function of wavelength and temperatures. Typically, photoinjectors use a fourth harmonic of a solid state laser (257-266 nm), and because of that the MTEs of photoelectronsgenerated from copper cathodes are often above the 100 meV level. With cryocooling and laser wavelength tuning MTEs can be lowered by one order of magnitude (reproduced from Ref.\cite{48}).}}
\end{figure}

\section*{Conclusions}

Today the production of spin polarized electron beams relies on photocathode materials and structures based on the III-V technology. State-of-the art photocathode structures based on superlattices embedded in Fabry-Perot resonant cavity enabled by a distributed Bragg reflector allow for the production of electron beams with simultaneous spin polarization above 80\% and quantum efficiencies larger than 5\%. The delicate nature of the surface layers needed to produce the required Negative Electron Affinity conditions represent the main obstacle for the production of high average current beams and operation of these photocathodes in RF electron guns. III-Nitrides materials showed to be more robust and are technologically ready to be studied as alternative to III-Vs for the production of spin polarized electron beams. Other classes of semiconductor materials belonging to the II-VI group have properties indicating their potential but it's unknown if they can be activated to Negative Electron Affinity. Magnetic structures not requiring any surface layer to enable the photoemission like spin filters and half metal might also be part of new directions worth exploring.

\section*{Acknowledgments}
The preparation of the manuscript was supported by the U.S. Department of Energy under grant DE-SC0012704.

\nolinenumbers

\bibliography{library}

\bibliographystyle{ieeetr}

\end{document}